# Cool dwarfs in wide multiple systems

## Paper I: Two mid-M dwarfs in a loosely-bound common-proper-motion pair

*By José A. Caballero*
*Centro de Astrobiología (CSIC-INTA), Madrid*

This is the first of a series of works devoted to investigate cool dwarfs in wide multiple systems. Here, I present Koenigstuhl 4 A and B, two bright, intermediate M dwarfs with a common high proper-motion and separated by 299 arcsec. At the most probable distance of the system, 19 pc, the projected physical separation is 5700 AU, which makes Koenigstuhl 4 AB to be one of the least bound binary systems with late-type components found to date. I also associate the primary with a *ROSAT* X-ray source for the first time.

*Introduction to the series 'Cool dwarfs in wide multiple systems'*

M-type dwarfs are the most abundant stars in the Universe. They provide us information on a wide range of topics, from formation of dust in their cool atmospheres ($T_{\rm eff} \sim$ 4000–2200 K)[1-3], through magnetic activity originated in their completely convective interiors[4-6], shape of the mass function down to the sub-stellar boundary[7-9], to possible existence of exoplanets in habitable zones that can be detected with current technology[10-12]. Only a few hundred cooler objects, with spectral types L ($T_{\rm eff} \sim$ 2200–1300 K) and T ($T_{\rm eff} \sim$ 1300–700 K)[13-15], have been catalogued to date. However, the characterization of such late-type stars and brown dwarfs is in general difficult because of their intrinsic faintness, especially at visible wavelengths.

Many known cool and ultracool (spectral type > M5V) objects are components of resolved physical systems[16-18]. When they are proper-motion companions to bright solar-like stars, their characterization is enormously eased: the heliocentric distance may be known (from *Hipparcos* parallax measurements), as well as their metallicities (which are very difficult to study in cool dwarfs)[19-21] and radial velocities (from which their kinematics in the Galaxy can be investigated)[22-24]. While most of the searches for cool and ultracool companions to nearby stars focus at close angular separations (of a few arcseconds), which require the use of high-resolution imaging techniques (such as adaptive optics) or facilities (such as the *Hubble Space Telescope*)[25-27], some benchmark late-type dwarfs are located at much wider separations (of several arcminutes)[28-30]. Moreover, the closest star to the Sun is a late-type dwarf at over two degrees from a solar-like binary system: Proxima Centauri (M5.5Ve)[31-33]. At the heliocentric distance of the α Centauri system, that angular separation translates into a physical separation of 12000 AU. This value contrasts with the tendency of equal-mass cool and ultracool binaries to be very tight[34-36], with physical separations rarely larger than 30 AU. However, there are a few examples of common proper-motion pairs with very low total masses that have separations of over 1000 AU and that represent a challenge for star-formation scenarios[37-39].

In this paper, I present the first item of a series of works devoted to investigate cool dwarfs in binary systems. They can be either newly discovered late-type wide companions to *Hipparcos* stars, poorly known equal-mass systems with very wide physical separations, or known systems with cool and ultracool components that require further characterization. I discover, recover, identify or notice the lack of information on the systems presented in this series while I accomplish other different programmes, mainly using the *Aladin* sky atlas of the virtual observatory[40], either as a single author or within a collaboration. In a sense, many of my targets are examples of serendipitous discoveries or by-products of large surveys.

*Koenigstuhl 4 AB: two mid-M dwarfs in a loosely-bound common proper-motion pair*

In this first item of the series, I present the serendipitous identification of two late-type stars that have similar high proper-motions and are separated by about 5 arcmin. They have been tabulated only by Luyten[41] (*NLTT*) and Lépine & Shara[42] (*LSPM*), and have never been presented as a bound binary system. The authors provided only their coordinates, approximate proper-motions and photographic *BRI* (in the visible) and *JHK$_s$* (in the near infrared) magnitudes. Besides, Luyten estimated "k-m" and "m" spectral types for the bright and faint component, respectively, from visible photometry.

In Table I, I compile names, coordinates and magnitudes from a number of sources (2MASS: Two-Micron All-Sky Survey[43]; CMC14: Carlsberg Meridian Catalogue 14[44]; USNO-B1.0: US Naval Observatory-B[45]; and references above) for the two components in the system, NLTT 6496 and NLTT 6491. Hereafter, I refer to them as Koenigstuhl 4 A and Koenigstuhl 4 B, respectively. This is a continuation of the *Koenigstuhl* nomenclature introduced by Caballero[37], in which new systems were designated with the name of a mountain near Heidelberg, Germany.

TABLE I
*Basic data of Koenigstuhl 4 A and B*

| Datum | A | B | Origin |
|---|---|---|---|
| NLTT | 6496 | 6491 | NLTT |
| LSPM | J0156+3033 | J0156+3028 | LSPM |
| α J2000 | 01 56 45.71 | 01 56 41.48 | 2MASS |
| δ J2000 | +30 33 28.8 | +30 28 48.9 | 2MASS |
| $B_J$ (mag) | 16.0 | 17.7 | NLTT |
| $r'$ (mag) | 14.449 | 16.318 | CMC14 |
| $R_F$ (mag) | 14.1 | 15.9 | NLTT |
| $I_N$ (mag) | 11.75 | 13.32 | USNO-B1.0 |
| $J$ (mag) | 10.323 | 11.917 | 2MASS |
| $H$ (mag) | 9.718 | 11.305 | 2MASS |
| $K_s$ (mag) | 9.449 | 11.029 | 2MASS |
| Sp. Type | M4.5±0.5V: | M6.5±0.5V: | *This work* |
| $M$ ($M_{sol}$) | $0.22^{+0.03}_{-0.04}$ | 0.12±0.01 | *This work* |

I identified the Koenigstuhl 4 system as a promising low-mass proper-motion system while searching for new and poorly-investigated late-M dwarfs bright enough to be suitable targets of radial-velocity searches for exoplanets. As illustrated in Fig. 1 and shown in Table II, the catalogued proper motions of Koenigstuhl 4 A and B are very similar, although not identical. Since the NLTT and LSPM proper-motion error bars are not tabulated, the USNO-B1.0 ones may be underestimated and the PPMXL (Positions and Proper Motions-Extended Large[46]) proper motions are marginally consistent only within 1σ uncertainties, a dedicated astrometric study was needed confidently to discard or confirm the common proper-motion of A and B.

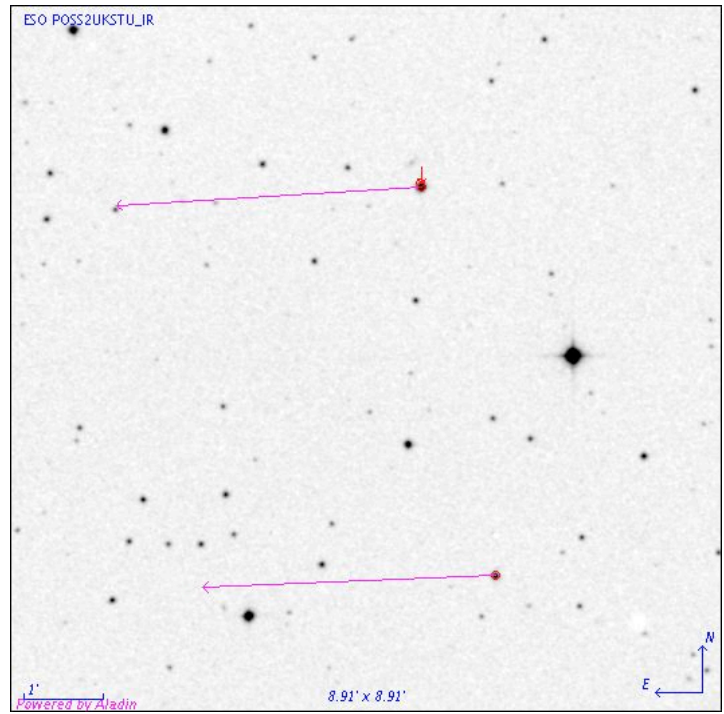

FIG. 1
UKST $I_N$-band image provided by *The Digitized Sky Survey* from the European Southern Observatory at Garching constructed with *Aladin* showing Koenigstuhl 4 A (top) and B (bottom). The long arrows indicate the LSPM proper motions as tabulated by *Simbad*. The primary is marked with a short arrow.

I followed the procedure shown by Caballero[47] to measure the proper motions of Koenigstuhl 4 A and B. I used eight (A) and seven (B) astrometric epochs (Table III) taken from 2MASS, CMC14, the SuperCOSMOS[48] digitizations of the photographic plates from the *United Kingdom Schmidt Telescope* (*UKST*) and first epoch of the Palomar Observatory Sky Survey (POSSI), the *Third US Naval*

*Observatory Astrograph Catalog* (*UCAC3*)[49], and the *Hubble Space Telescope Guide Star Catalogue*, version 1.2 (GSC1.2)[50]. The first and last epochs were separated by 51 years. Koenigstuhl 4 B was not catalogued by GSC1.2 (epoch 1983 September 08).

TABLE II
*Proper motions of Koenigstuhl 4 A and B*

| Origin | A | | B | |
|---|---|---|---|---|
| | $\mu_\alpha \cos\delta$ (mas a$^{-1}$) | $\mu_\delta$ (mas a$^{-1}$) | $\mu_\alpha \cos\delta$ (mas a$^{-1}$) | $\mu_\delta$ (mas a$^{-1}$) |
| NLTT | +230 | –16 | +259 | –41 |
| USNO-B1.0 | +220±2 | –10±1 | +210±1 | –10±1 |
| LSPM | +229 | –13 | +220 | –8 |
| PPMXL | +219±5 | –12±5 | +208±4 | –12±4 |
| *This work* | +213.0±1.1 | –15.1±0.4 | +205.8±1.2 | –15.5±0.5 |

The proper-motions measured by me (last row of Table II) resemble each other more than in the case of *NLTT*, USNO-B1.0, LSPM or PPMXL. The resemblance led to a definitive test, which was to measure the constancy with time of the angular separation (ρ) and position angle (θ) between Koenigstuhl 4 A and B. That test is more reliable than measuring the proper motions of the components separately because it is not affected by the parallax (indeed, the parallactic wobbling is the main contributor to the 'large' error bars of 1.1–1.2 mas a$^{-1}$ in right-ascension proper motion).

Table III summarises the ρ and θ measurements for the seven available astrometric epochs. The UCAC3 and CMC14 epochs of A and B were not simultaneous, but separated by 7 and 62 days, respectively, so I took the average date for the computation. In spite of that source of error, the standard deviations of the mean angular separation and position angle in the 51-year interval kept rather small, of only 90 mas and 0.05 deg, respectively, which confidently confirmed the common proper motion of Koenigstuhl 4 A and B.

TABLE III
*Astrometric observations of the Koenigstuhl 4 AB system*

| Epoch | ρ (arcsec) | θ (deg) | Origin |
|---|---|---|---|
| 1951 Nov. 05 | 298.9±0.3 | 190.4±0.2 | POSSI |
| 1991 Sep. 17 | 298.9±0.3 | 190.5±0.2 | POSSII Red |
| 1993 Sep. 15 | 299.0±0.3 | 190.5±0.2 | POSSII Blue |
| 1995 Oct. 15 | 299.1±0.3 | 190.5±0.2 | POSSII Infrared |
| 1999 Nov. 05 | 298.96±0.10 | 190.5±0.2 | 2MASS |
| 2001 Oct. 05 | 298.85±0.07 | 190.4±0.2 | UCAC3 |
| 2003 Dec. 06 | 299.1±0.2 | 190.6±0.2 | CMC14 |

Once the common proper motion was confirmed, the system needed to be characterized in detail. The basic data of Koenigstuhl 4AB as a system are summarized in Table IV. By comparing the $r'$–$J$ and $J$–$K_s$ colours measured for Koenigstuhl 4 A and B with those of known M dwarfs with Sloan Digital Sky Survey (SDSS; $u'g'r'i'z'$) and 2MASS ($JHK_s$) photometry from West *et al.*[51] (CMC14 $r'$ magnitude is in the SDSS system) and the magnitude difference between the two components, I concluded that the only possible spectral type combinations are M4.0V+M6.0V, M4.5V+M6.5V or M5.0V+M7.0V (see the bottom of Table I). Assuming the intermediate combination as the most probable one and spectral type uncertainties of 0.5 sub-types, and using the spectral type-*J*-band absolute-magnitude relationship given by Caballero *et al.*[52], I derived a conservative distance of $19^{+6}_{-4}$ pc. From the $M_J$ magnitudes and the NextGen models[53] for ages older than 1 Ga, I estimated the masses of Koenigstuhl 4 A and B at around 0.22 and 0.12 $M_{sol}$, which gave the system a total mass of only around 0.34 $M_{sol}$ (Table IV).

The spectral-type estimation is supported by the existence of a nearby *ROSAT* All-Sky Survey X-ray source, 1RXS J015645.8+303332 (count rate $CRT$ = 4.4±1.5 ks$^{-1}$)[54], at only 3.3 arcsec to Koenigstuhl 4 A. Because of its star-like hardness ratios ($HR_1$ = –0.5±0.3, $HR_2$ = –0.7±1.4) and the large incidence of X-ray activity among mid-M dwarfs[55-57], I propose that the origin of the 1RXS source is Koenigstuhl 4 A itself.

TABLE IV
*Basic data of the Koenigstuhl 4 AB system*

| | |
|---|---|
| <$\rho$> (arcsec) | 298.97±0.09 |
| <$\theta$> (deg) | 190.49±0.05 |
| $d$ (pc) | $19^{+6}_{-4}$ |
| $s$ (AU) | $5700^{+1800}_{-1200}$ |
| $M_{total}$ ($M_{sol}$) | $0.34^{+0.04}_{-0.05}$ |
| –$U^*_g$ (J) | 7.7–8.2 $10^{33}$ |

The projected physical separation $s$ of around 5700 AU makes Koenigstuhl 4 AB one of the widest systems with $M_{total}$ < 0.4 $M_{sol}$. A very wide binary is synonymous with a weakly bound system in the process of being disrupted by the gravitational potential of the Galaxy[58-60]. As a proxy to the gravitational potential energy $U_g$, I computed $U^*_g$ = –$G$ $M_1$ $M_2/s$ for Koenigstuhl 4 AB, which is one of the lowest binding energies known. I provide an interval of –$U_g$ in Table IV because it peaks at the most probable values (M4.5V+M6.5V at $d$ = 19 pc); cooler spectral types imply a smaller physical separation (*i.e.*, a shorter distance) but less total mass while warmer spectral types imply a larger total mass but a wider physical separation (*i.e.*, a longer distance).

Koenigstuhl 4 AB falls in a binding-energy/total-mass diagram (*e.g.*, Fig. 2 in

Caballero[60]) very close to the only representatives of the rare class of very wide ($s > 1000$ AU), very low-mass ($M_1+M_2 < 0.2$ $M_{sol}$), equal-mass ($M_2/M_1 \sim 1$) binaries: Koenigstuhl 1 AB ($s \sim 1800$ AU)[37], 2M0126–50 AB ($s \sim 5200$ AU)[38] and 2M1258+40 AB ($s \sim 6700$ AU)[39], which are the least-bound multiple systems with low-mass components known to date with the exception of the very young brown-dwarf plus exoplanet system 2M1207–39 AB[61]. The larger total mass of Koenigstuhl 4 AB is compensated by its wider physical separation, which makes the system a new challenge for low-mass star-formation scenarios and studies of dynamical encounters in the Galactic disc. Although there is an obvious lack of data on the system (*i.e.*, effective temperatures, surface gravities, activity indicators, radial velocities, accurate photometry in the optical) one would venture to guess that Koenigstuhl 4 AB is in the process of disruption.

Low-resolution optical spectroscopy for accurate spectral-type determination and study of the Hα emission (a signpost of activity) of Koenigstuhl 4 A and B is planned for the near future.